\documentclass[5p]{elsarticle}

\usepackage{lineno,hyperref,amssymb,amsmath,amsthm}
\modulolinenumbers[5]

\begin{document}

\begin{frontmatter}

\title{New Ekpyrotic Quantum Cosmology}

\author{Jean-Luc Lehners}
\address{Max-Planck-Institute for Gravitational Physics (Albert-Einstein-Institute) \\ Am M\"{u}hlenberg 1, D-14476 Potsdam, Germany}
\ead{jlehners@aei.mpg.de}

\begin{abstract}
Ekpyrotic instantons describe the emergence of classical contracting universes out of the no-boundary quantum state. However, up to now these instantons ended in a big crunch singularity. We remedy this by adding a higher-derivative term, allowing a ghost condensate to form. This causes a smooth, non-singular bounce from the contracting phase into an expanding, kinetic-dominated phase. Remarkably, and although there is a non-trivial evolution during the bounce, the wavefunction of the universe is ``classical'' in a WKB sense just as much after the bounce as before. These new non-singular instantons can thus form the basis for a fully non-singular and calculable ekpyrotic history of the universe, from creation until now.
\end{abstract}


\end{frontmatter}


\paragraph{Motivation and aims} The goal of cosmology is to explain the (broad) evolution of the universe, in its full spatial and temporal extent. This programme is plagued by the occurrence of singularities and/or infinities, where we lose control over the theory and hence require the additional input of initial or boundary conditions. In other words, as long as singularities are present, there are necessarily \emph{ad hoc} elements in any cosmological model, implying that all results are based on assumptions that lie outside of the theory. Inflation is typically preceded by a singularity \cite{Borde:2001nh}, in much the same way as the old hot big bang model is preceded by the big bang singularity \cite{Penrose:1964wq,Hawking:1969sw}. This situation can be improved in quantum cosmology where one can formulate a theory of initial conditions, such as the Hartle-Hawking no-boundary proposal that we will focus on here \cite{Hartle:1983ai}\footnote{Recent work on the Hartle-Hawking proposal includes \cite{Hartle:2015bna,Chen:2015ria}.}. Inflationary instantons then render the beginning of the universe, along with its initial conditions, non-singular and calculable \cite{Hartle:2008ng}. However, typically inflation is thought to lead to eternal inflation and the multiverse \cite{Steinhardt:1982kg,Vilenkin:1983xq}, which brings along its own elements of incalculability due to the infinities that arise from the continued production of new regions of the universe with different physical properties \cite{Ijjas:2013vea,Guth:2013sya,Linde:2014nna,Ijjas:2014nta}. Here we try to look at these issues in an alternative framework, namely that of ekpyrotic models \cite{Khoury:2001wf,Lehners:2008vx}.

These models do not amplify large quantum fluctuations and do not lead to a run-away behaviour like eternal inflation \cite{Johnson:2011aa}. Rather, the universe is rendered smooth during a slowly contracting ekpyrotic phase, which has the effect of homogenising the universe over large regions \cite{Erickson:2003zm}. It was recently shown that such universes can also be described in quantum cosmology, via \emph{ekpyrotic instantons} \cite{Battarra:2014xoa,Battarra:2014kga,Lehners:2015sia}. However, all of the instantons presented to date have the drawback of ending in a big crunch singularity, so that one has to hope for a full quantum gravity resolution of the singularity. In that case, one may wonder what good it is to describe a previous phase of the universe in semi-classical quantum gravity, when there is again an unknowable aspect linking this phase to the present expanding phase. Classically, this situation was addressed within the framework of \emph{new ekpyrotic} cosmology \cite{Buchbinder:2007ad}, where the original brane collision bounce of the ekpyrotic model was replaced by a classically non-singular bounce. This leads to a calculable cosmology, where also the transfer of fluctuations across the bounce is unambiguous \cite{Battarra:2014tga}. Here we combine all of the ideas above to find instantons that describe the emergence of an ekpyrotic universe followed by a non-singular bounce into the present expanding phase. These solutions provide the basis for a semi-classical, non-singular and fully calculable history of the universe.

\begin{figure}[h] 
\begin{center}
\includegraphics[width=0.35\textwidth]{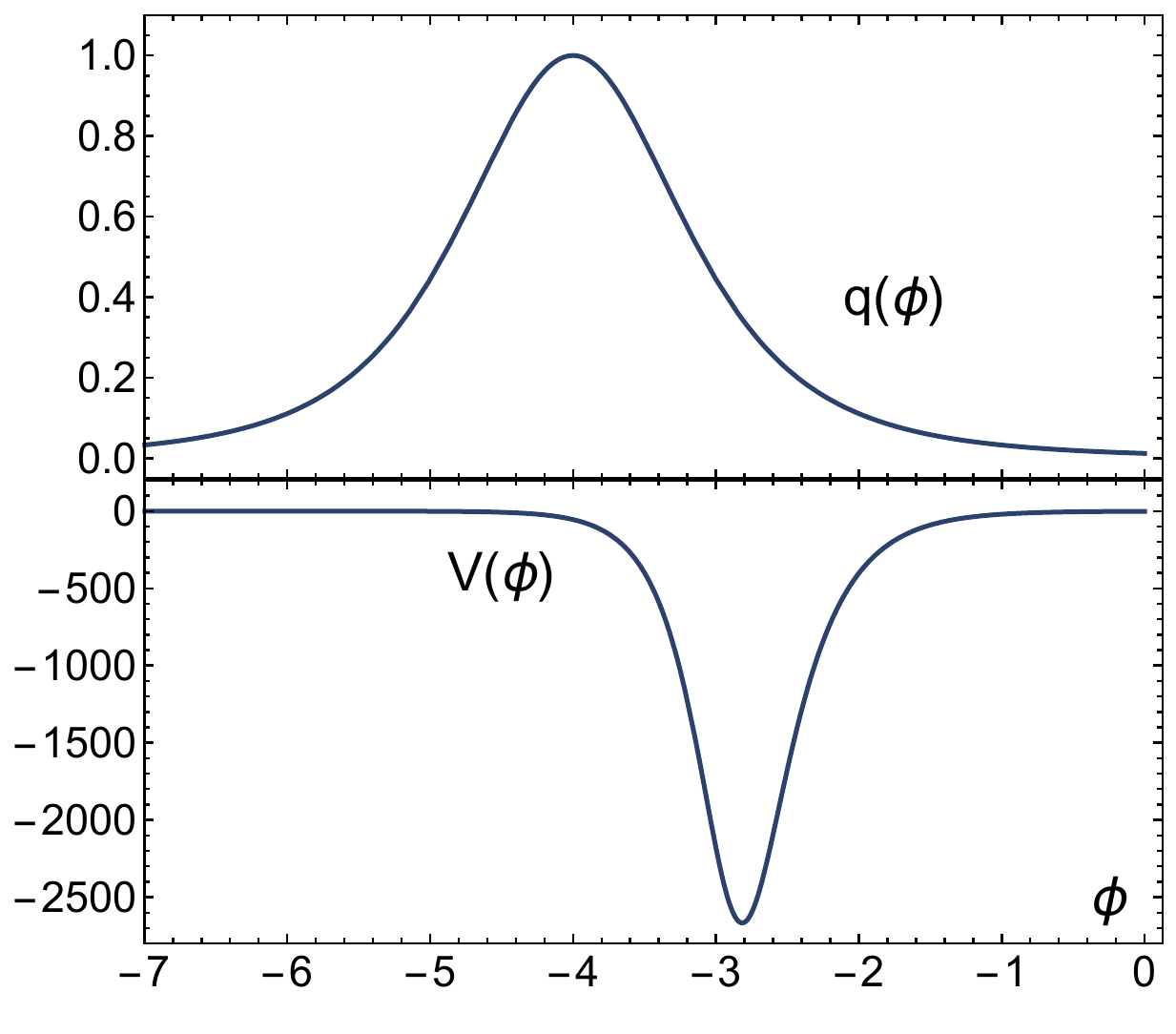}
\caption{The kinetic function $q(\phi)$ (top panel) governing the higher-derivative $X^2$ term, and the potential $V(\phi)$ with $V_0=1$ (bottom panel). The scalar field starts on the right; the ekpyrotic phase lasts until the potential starts turning off, when the bounce phase begins.}
\label{FigQinPot}
\end{center}
\end{figure}

\paragraph{Setup}
In order to describe both an ekpyrotic phase and a bounce, we adopt a model of a scalar field coupled to gravity, and where the kinetic term can contain higher-derivative terms that are functions of the ordinary kinetic term $X\equiv -\frac{1}{2}(\partial\phi)^2,$ i.e. we consider the model defined via the action (in natural units)
\begin{equation}
S = \int \sqrt{-g} \left[ \frac{R}{2} + P(X,\phi) \right]\,.
\end{equation}
We will specialise to the case where $P(X,\phi) = k(\phi)X + q(\phi)X^2 - V(\phi).$ The inclusion of the $X^2$ term is enough to cause a ghost condensate bounce, as long as the ordinary kinetic term switches sign \cite{Buchbinder:2007ad,Creminelli:2007aq}. Choosing the free functions in the action to take the form (see Fig. \ref{FigQinPot}) 
\begin{eqnarray}
k(\phi) &=& 1 - \frac{2}{\left(1 + \frac{1}{2}(\phi + 4)^2\right)^2}\\ q(\phi) &=& \frac{1}{\left(1 + \frac{1}{2}(\phi + 4)^2\right)^2}\\ V(\phi) &=& - \frac{V_0}{\left( e^{3\phi} + e^{-4(\phi+5)}\right)}
\end{eqnarray}
ensures that the potential is of ekpyrotic form for $\phi \gtrsim -3$ and that the kinetic term switches sign while the higher-derivative term is turned on. Then a bounce can occur after the ekpyrotic phase has come to an end, i.e. for $\phi \lesssim -3$ \cite{Cai:2012va,Koehn:2013upa}. These functions are chosen here for convenience -- the rough shapes are important, but there is a significant amount of freedom in the specific functional form. 

\begin{figure}[h] 
\begin{center}
\includegraphics[width=0.35\textwidth]{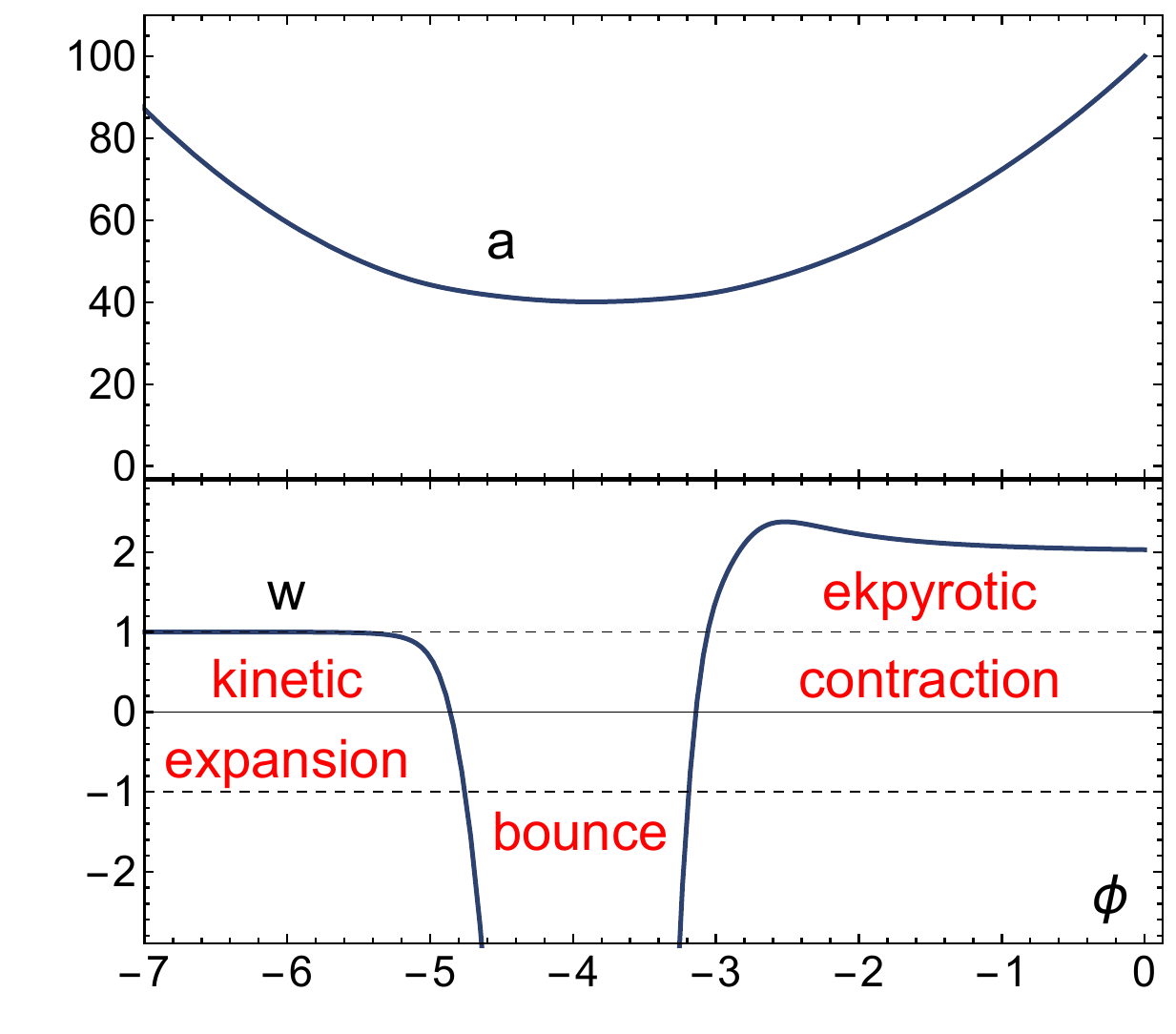}
\caption{The scale factor $a$ (top panel) and the equation of state $w$ (bottom panel) for a non-singular bounce solution. It consists of an ekpyrotic phase (for $\phi \gtrsim -3$), a bounce ($-5 \lesssim \phi \lesssim -3$) and a kinetic dominated expanding phase ($\phi \lesssim -5$).}
\label{FigSolPlot}
\end{center}
\end{figure}

In the present study we will specialise to line elements of Robertson-Walker form with closed spatial sections
\begin{equation}
\mathrm{d}s^2 = N^2 \mathrm{d}\lambda^2 + a^2(\lambda) \mathrm{d}\Omega_3^2\,,
\end{equation}
where the lapse function $N$ is real in the Euclidean case and pure imaginary for Lorentzian signature universes, and where $\mathrm{d}\Omega_3^2$ is the metric on the unit three-sphere. Writing $\mathrm{d}\tau = N \mathrm{d}\lambda$ and denoting a derivative w.r.t the generally complex time $\tau$ by a prime, the equations of motion become
\begin{eqnarray}
3 \frac{a^{\prime 2}}{a^2} - \frac{3}{a^2} \!\!\! &=& \!\!\! \frac{1}{2}k\phi^{\prime 2} - \frac{3}{4}q\phi^{\prime 4} - V\,, \\ - 2 \frac{a^{\prime\prime}}{a} - \frac{a^{\prime 2}}{a^2} + \frac{1}{a^2} \!\!\! &=& \!\!\! \frac{1}{2}k\phi^{\prime 2} - \frac{1}{4}q\phi^{\prime 4} + V\,, \\ (k-3q\phi^{\prime 2}) \phi^{\prime\prime} + (k_{,\phi} \!\!\!\! &-& \!\!\!\! q_{,\phi}\phi^{\prime 2})\phi^{\prime 2} + 3 \frac{a^\prime}{a}\phi^\prime (k - q \phi^{\prime 2}) \nonumber \\ &=& \!\!\! \frac{1}{2}k_{,\phi}\phi^{\prime 2} - \frac{1}{4}q_{,\phi}\phi^{\prime 4} + V_{,\phi} \,.
\end{eqnarray}
The Euclidean action $S_E = - iS$ then reduces on-shell to
\begin{equation} \label{Eaction}
S_E = \pi^2 \int \mathrm{d}\tau \left[ - 12 a + 4 a^3 V + a^3 q \phi^{\prime 4} \right]\,.
\end{equation} 
A classical (Lorentzian) solution of the field equations, starting in the ekpyrotic phase and proceeding via a smooth bounce into a kinetic dominated expanding phase is shown in Fig. \ref{FigSolPlot} (note that the evolution is towards more negative values of the scalar field $\phi$).

The no-boundary wavefunction is formulated in the Euclidean path integral formalism (Ref. \cite{Battarra:2014kga} contains a review and uses the same notation as here), with the result that in the semi-classical approximation, which we will use throughout, the wavefunction is of the form
\begin{equation}
\Psi(b,\chi) = e^{-S_E(a,\phi)}\,,
\end{equation}
where $S_E(a,\phi)$ is the action of a \emph{complex} solution $a(\tau), \phi(\tau)$ satisfying the equations of motion above with the boundary conditions
\begin{itemize}
\item $a(\tau=0) = 0, \, a^\prime(0)=1, \, \phi^\prime(0)=0$ -- this is the \emph{no boundary} condition, which ensures that the instanton is regular and without boundary at the so-called South Pole $\tau = 0.$
\item The scale factor and scalar field take the specified values $a(\tau_f) = b,\, \phi(\tau_f) = \chi,$ with $b$ and $\chi$ being real numbers, at a final time $\tau_f.$
\end{itemize}
Both the (generally complex) scalar field value at the South Pole $\phi_{SP}$ as well as the final time $\tau_f$ must be determined such that the conditions above are satisfied. Note that in the present approach it is precisely the extension to complex field values that allows for quantum effects to be taken into account.

\begin{figure}[h] 
\begin{center}
\includegraphics[width=0.35\textwidth]{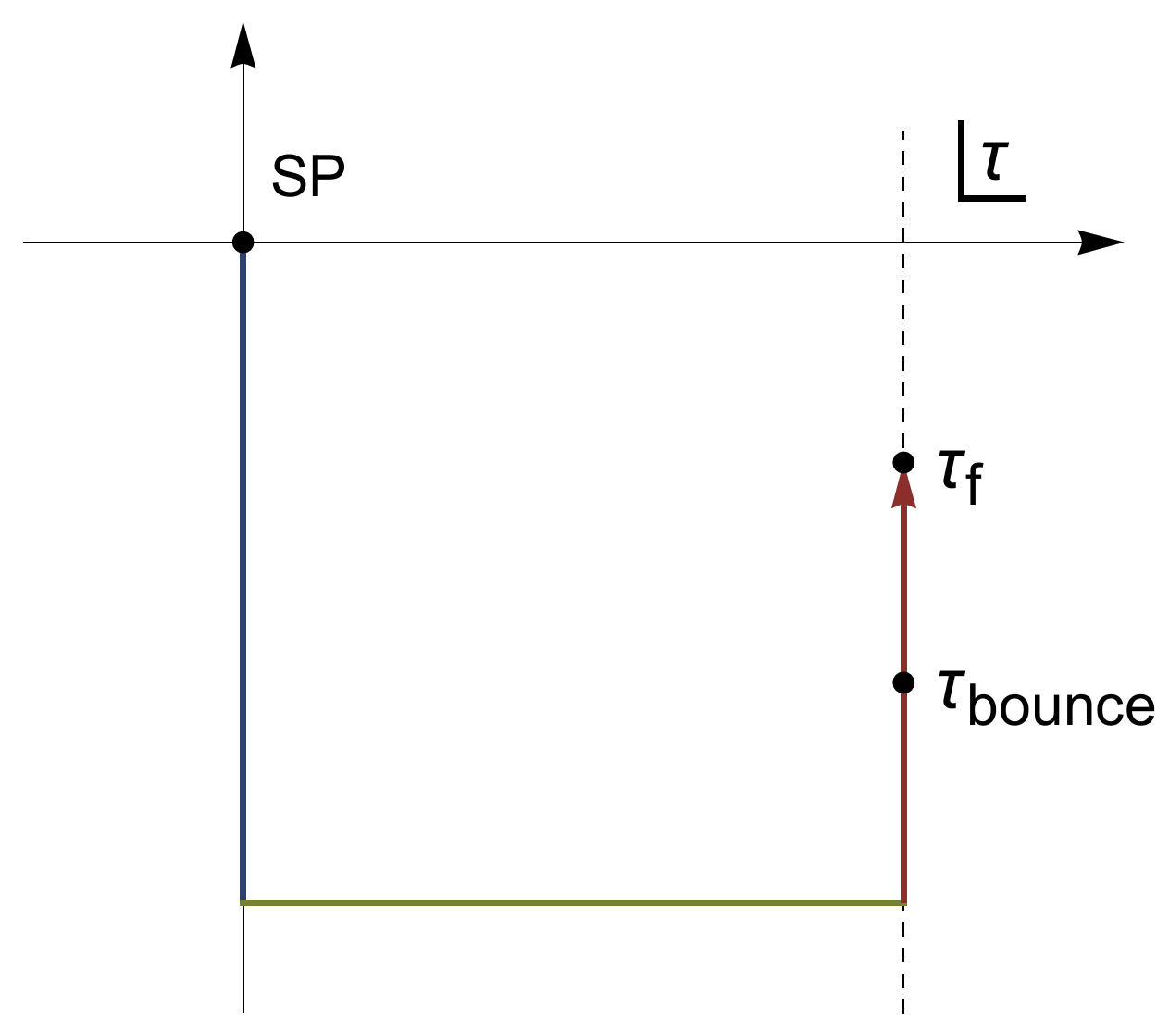}
\caption{In evaluating instantons we choose the contour depicted above with three consecutive segments: down from the South Pole at $\tau=0$ with lapse function $N=-i,$ then across with $N=1$ and finally up along a Lorentzian direction $N=i.$ The contour is chosen to run sufficiently far down so as to reach the end of the ekpyrotic phase and the bounce on the final segment.}
\label{Figcontour}
\end{center}
\end{figure}

With the prescription above, the Euclidean action \eqref{Eaction} can be viewed as a contour integral in the complex time plane, from $\tau=0$ to a final time $\tau_f.$ We will consider contours such as the one depicted in Fig. \ref{Figcontour}, i.e. we take the contour down along the imaginary axis, then across the real $\tau$ axis and finally up along the imaginary $\tau$ direction until reaching $\tau_f.$ Along the last segment of the contour the lapse function is pure imaginary and hence the metric is Lorentzian as long as the scale factor and scalar field are real. Typically they are only simultaneously real exactly at the final time $\tau_f,$ but for increasingly classical universes we expect that the instantons change less and less as the wavefunction $\Psi(b,\chi)$ is evaluated along a classical history $b(\lambda),\chi(\lambda).$ Looking at Eq. \eqref{Eaction} and keeping in mind that along the last segment $\mathrm{d}\tau = i \mathrm{d}t$ with $t \in \mathbb{R},$ this then implies that (while progressing along the family of instantons corresponding to the classical history of interest) the real part of the Euclidean action $S_E^R$ ought to change less and less compared to the imaginary part $S_E^I$. A useful quantitative criterion for classicality stems from the analogy with the Wentzel-Kramers-Brillouin (WKB) method in quantum mechanics: for a classical universe we expect the amplitude of the wavefunction to vary slowly compared to the variation of its phase, i.e. we would want
\begin{equation} \label{WKB}
\frac{\partial_b S_E^R}{\partial_b S_E^I} \ll 1,\quad \frac{\partial_\chi S_E^R}{\partial_\chi S_E^I} \ll 1\,.
\end{equation}
In this case, one may also associate a relative (unnormalised) probability $e^{-2 S_E^R}$ to the corresponding history.

\begin{figure}[h] 
\begin{center}
\includegraphics[width=0.4\textwidth]{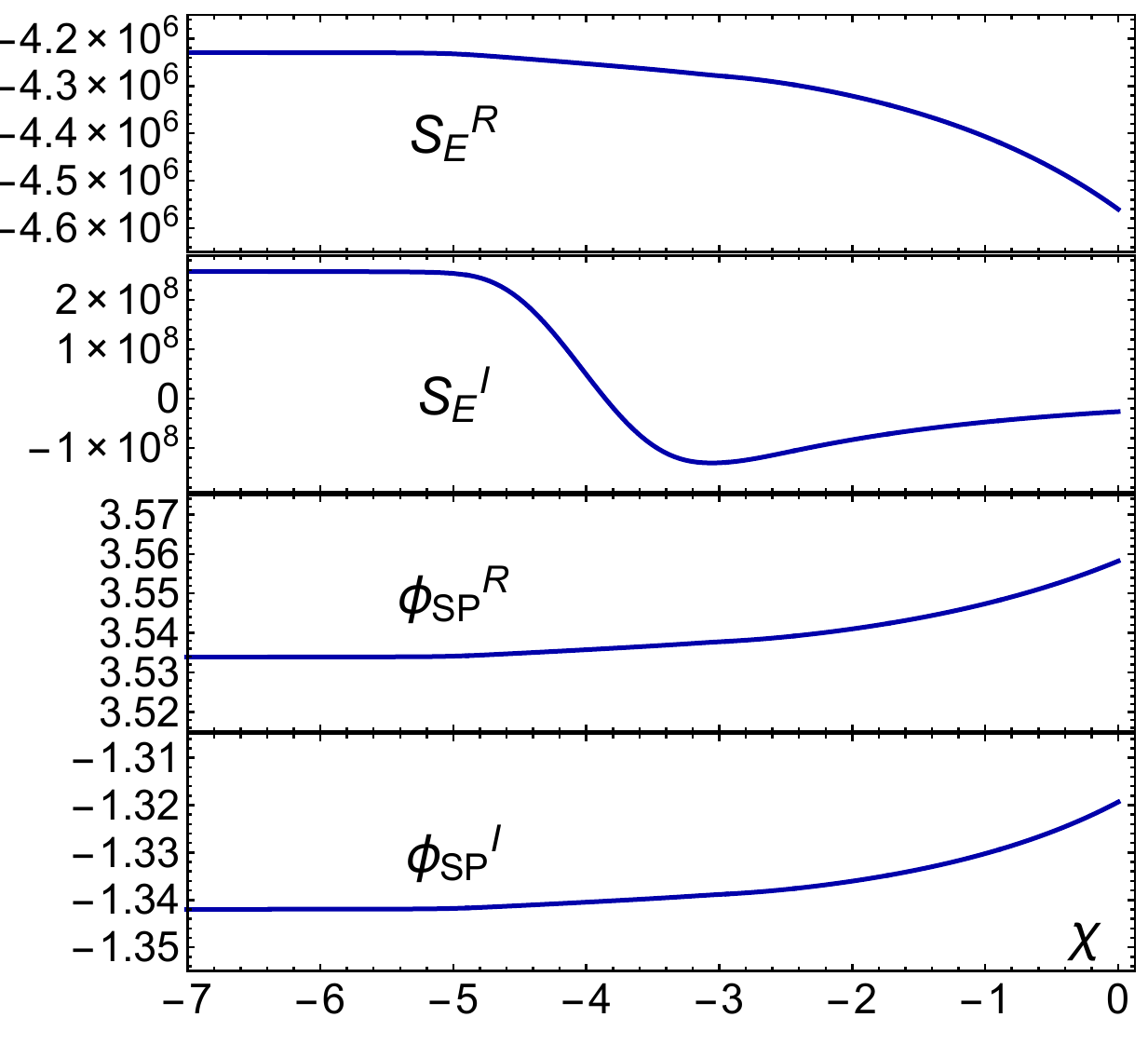}
\caption{The values of the Euclidean action and of the scalar field at the South Pole of the instanton (i.e. at the ``no boundary'' point), both real and imaginary parts, for the family of instantons corresponding to the classical history \eqref{history}. The evolution is smooth across the bounce and the values of $\phi_{SP}$ and $S_E^R$ stabilise as the history progresses from $\chi=0$ to more negative values of $\chi.$}
\label{FigCombined}
\end{center}
\end{figure}

\paragraph{Numerical results}
For specificity we choose $V_0 = 10^{-6}$ and in order to facilitate the numerical evaluations, the theory is re-scaled according to
\begin{equation}
\bar{a} = V_0^{1/2} a\,,\quad \bar{\phi} = \phi\,,
\end{equation}
which implies that $k(\phi)$ is unchanged while $q(\phi)$ is effectively re-scaled to $\bar{q} = V_0 q$ and $\bar{V}=V_0^{-1} V.$ As a representative example, we choose a classical history specified by 
\begin{equation} \label{history}
\bar{a}(\lambda_i)=100\,, \, \bar{\phi}(\lambda_i) = 0\,, \, \bar{\phi},_{\bar\lambda}(\lambda_i) = -2.4555\,,
\end{equation}
which is also the solution shown in Fig. \ref{FigSolPlot}. We have evaluated the no-boundary wavefunction along this classical history, at $427$ roughly equally spaced values in $\chi$ between $0$ and $-7.$ The corresponding South Pole values of the scalar field are shown in Fig. \ref{FigCombined}. These values, as well as the real part of the Euclidean action (also shown in the figure), are seen to vary very little and are converging to asymptotic values, which is due (at least in part) to the attractor behaviour of the ekpyrotic phase \cite{Lehners:2015sia}. Crucially, the evolution through the bounce proceeds uneventfully, and the values of $\phi_{SP}$ and $S_E$ are smooth. This already demonstrates that there is no obstacle in incorporating a non-singular bounce into the wavefunction of the universe.

\begin{figure}[h] 
\begin{center}
\includegraphics[width=0.35\textwidth]{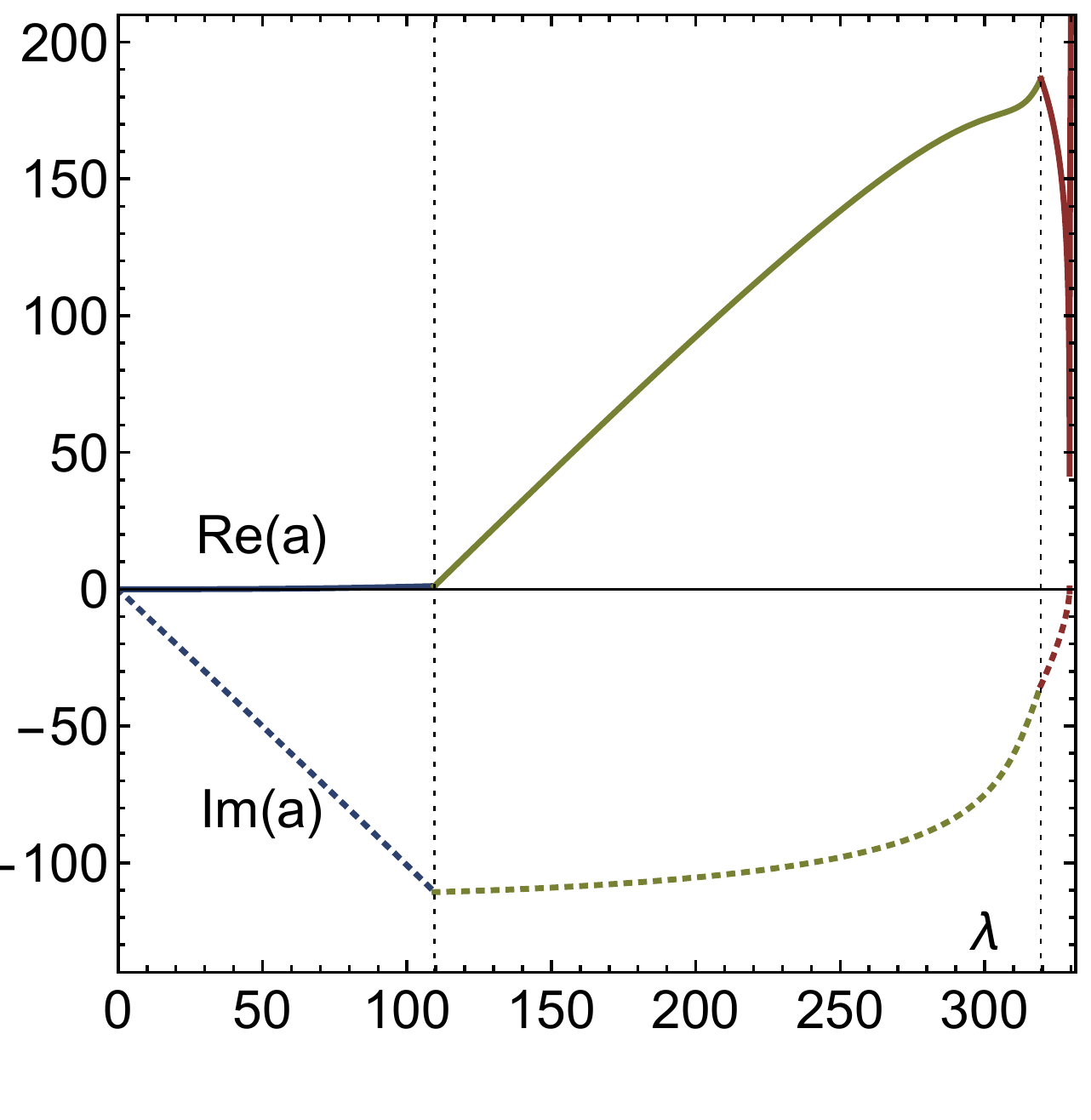}
\caption{The real and imaginary parts of the scale factor, for a representative instanton ($b=86.0,\,\chi=-6.97$).}
\label{FigaPlot}
\end{center}
\end{figure}

\begin{figure}[h] 
\begin{center}
\includegraphics[width=0.35\textwidth]{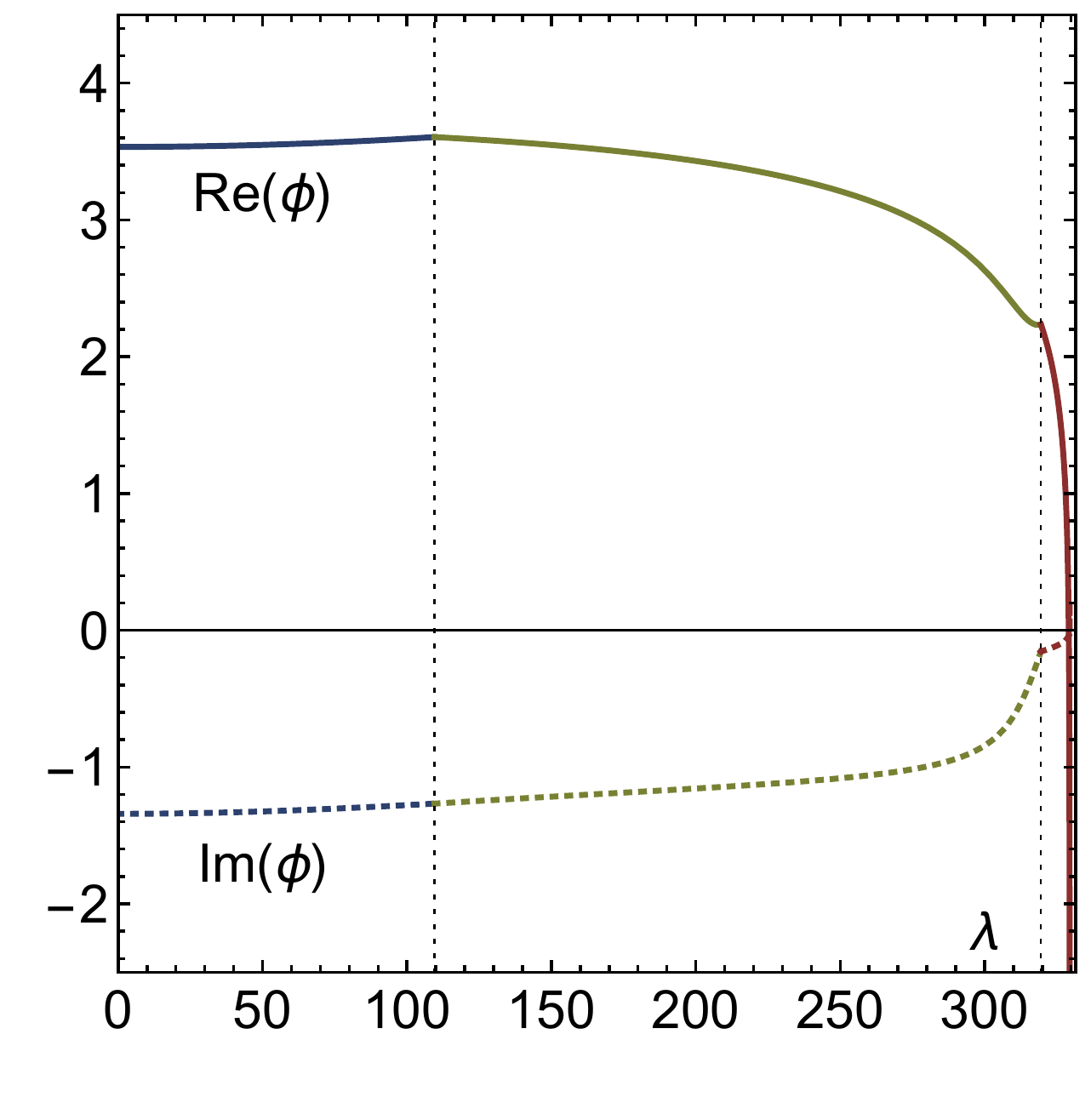}
\caption{The real and imaginary parts of the scalar field, for the same instanton ($b=86.0,\,\chi=-6.97$).}
\label{FigPhiPlot}
\end{center}
\end{figure}

\begin{figure}[h] 
\begin{center}
\includegraphics[width=0.45\textwidth]{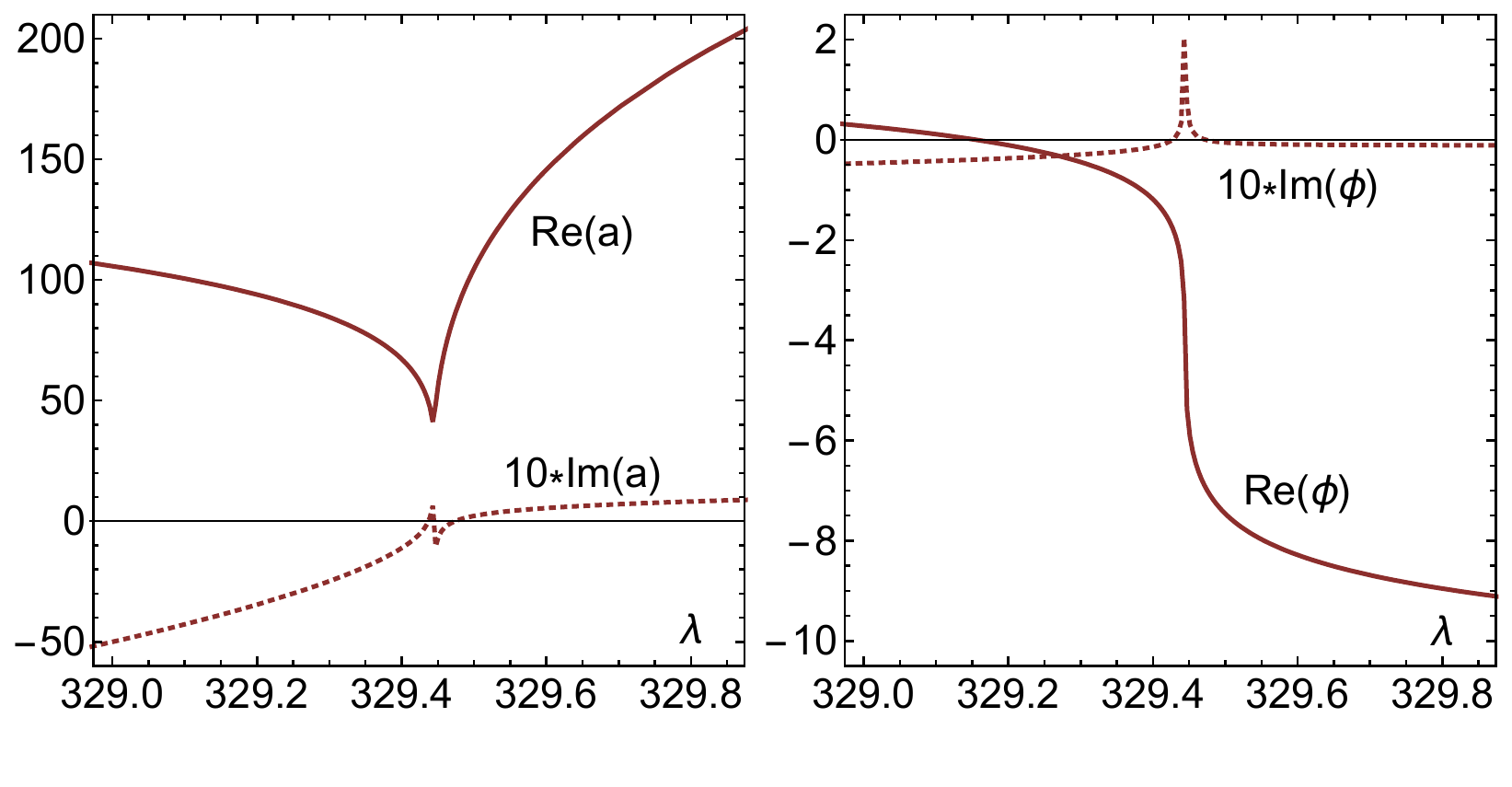}
\caption{A zoom-in of Figs. \ref{FigaPlot} and \ref{FigPhiPlot} along the final segment, showing the bounce region. Note that the bounce appears less smooth here than in Fig. \ref{FigSolPlot}, because that figure shows a parametric $(a,\phi)$ plot and the scalar field varies significantly across the bounce, as can be seen above. Here the imaginary parts have been amplified $10$ times to aid the visualisation. The scale factor and scalar field are only simultaneously exactly real at the specified values ($b=86.0,\,\chi=-6.97$), but the imaginary parts remain small afterwards, as the universe continues expanding.}
\label{Figzoom}
\end{center}
\end{figure}

A representative (late) instanton is shown in Figs. \ref{FigaPlot} and \ref{FigPhiPlot}, where the real and imaginary parts of the scale factor and scalar field are shown along a contour of the form described in Fig. \ref{Figcontour}. Up to the bounce this instanton has a shape that is similar to that of crunching ekpyrotic instantons \cite{Battarra:2014xoa}, as expected. Along the first segment (in blue), the instanton consists of a disk of Euclidean flat space. The second segment (in beige) is fully complex and interpolates to the third segment (in red) along which the ekpyrotic phase takes place. It is here that the universe becomes classical, as the imaginary parts of the fields are becoming small. However, unlike the previously studied ekpyrotic instantons of Refs. \cite{Battarra:2014xoa,Battarra:2014kga,Lehners:2015sia}, the universe does not end in a crunch, but rather the higher-derivative term becomes important after the end of the ekpyrotic phase and induces a non-singular bounce, as can better be seen from the zoom-ins in Fig. \ref{Figzoom}. After the bounce, the universe keeps expanding in a kinetic dominated phase. One might then imagine that subsequently the scalar field decays and fills the universe with radiation and matter.

\begin{figure}[h] 
\begin{center}
\includegraphics[width=0.35\textwidth]{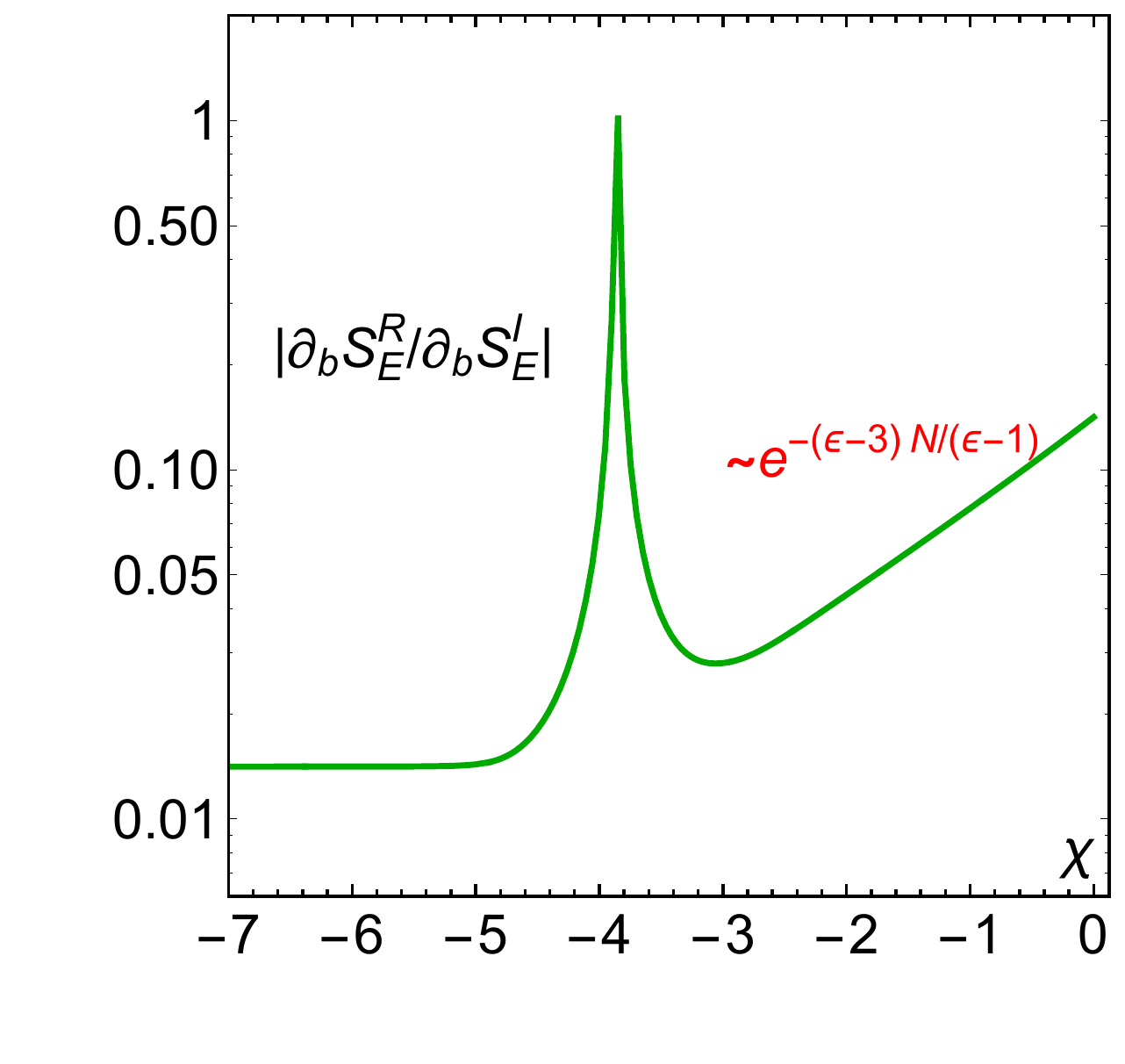}
\caption{The first of the WKB classicality conditions \eqref{WKB} along the classical bouncing history \eqref{history}. The evolution is from right to left, and a smaller value indicates a more classical wavefunction. This figure is discussed in detail in the main part of the text.}
\label{FigWKBb}
\end{center}
\end{figure}

\begin{figure}[h] 
\begin{center}
\includegraphics[width=0.35\textwidth]{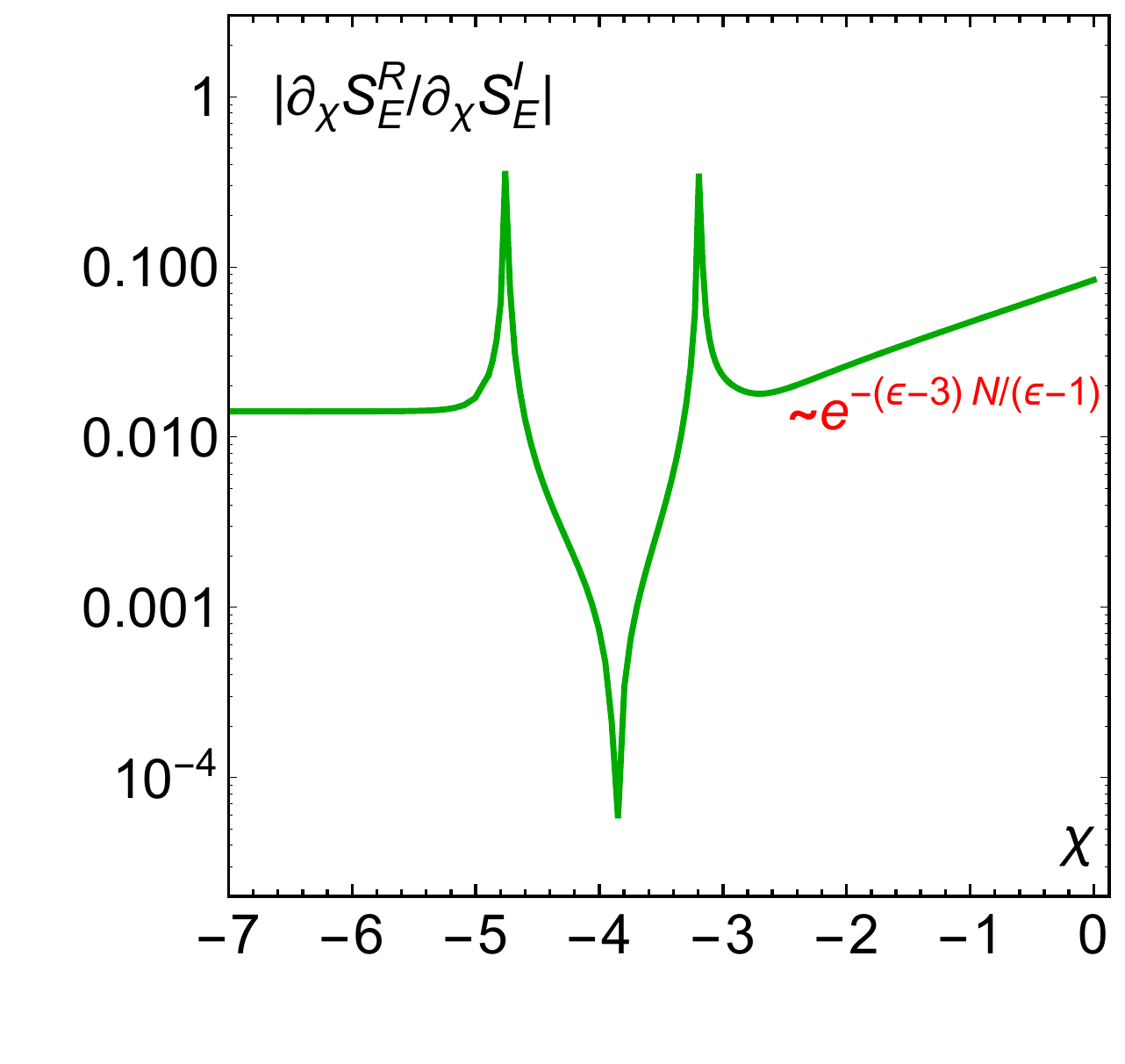}
\caption{The second of the WKB classicality conditions \eqref{WKB} along the classical bouncing history \eqref{history}. See the main part part of the text for a full discussion.}
\label{FigWKBchi}
\end{center}
\end{figure}

Of special interest are the WKB classicality conditions, which can tell us whether the universe is still classical after the bounce. This question is somewhat non-trivial, as the null energy condition is violated during the bounce phase (as the equation of state $w$ drops below $-1$). One may wonder whether the wavefunction remains classical under such extreme conditions. In order to settle this issue, one needs to evaluate the derivatives of the wavefunction w.r.t to $b$ and $\chi.$ These derivatives can be accurately approximated by finite difference estimates (see the previous works \cite{Battarra:2014kga,Lehners:2015sia}), obtained by evaluating the wavefunction for slightly shifted classical histories $\left(b+b\cdot \delta b,\chi\right)$ and $\left(b,\chi + \delta\chi\right)$ with $\delta b = \delta \chi = 10^{-5}.$ The results are shown in Figs. \ref{FigWKBb} and \ref{FigWKBchi}. On the right of both figures, during the ekpyrotic phase, the WKB conditions scale as $e^{-(\epsilon - 3)N/(\epsilon - 1)},$ where $\epsilon = \frac{3}{2}(1+w)$ is the fast-roll parameter of the ekpyrotic phase. This scaling, which applies when the equation of state is constant, was first derived in \cite{Battarra:2014kga} and discussed extensively in \cite{Lehners:2015sia}. 

New to this paper is the subsequent evolution during the bounce phase. A number of notable features are seen here: the first is that at isolated moments, the WKB conditions take on a cusp shape and blow up (an analytical treatment might very well be required to achieve a full understanding of this aspect). More specifically, the variation w.r.t. the scale factor $b$ momentarily blows up at the moment of the bounce (Fig. \ref{FigWKBb}). The fact that the Euclidean action itself evolves smoothly and remains finite all along (see Fig. \ref{FigCombined}) shows that in the ratio $\partial_b S_E^R/\partial_b S_E^I$ the denominator $\partial_b S_E^I$ becomes momentarily zero rather than $\partial_b S_E^R$ becoming infinite. In other words, the imaginary part of the Euclidean action momentarily becomes stationary with regard to $b$ at the bounce. Furthermore, the WKB condition w.r.t the scalar field blows up at two instants, which appear to be the times when the null energy condition is marginally satisfied, i.e. when $w=-1;$ compare Figs. \ref{FigWKBchi} and \ref{FigSolPlot}. At these moments the imaginary part of the Euclidean action becomes stationary with regard to $\chi.$ In addition, and in opposition to what happens in Fig. \ref{FigWKBb}, the combination $\partial_\chi S_E^R/\partial_\chi S_E^I$ goes to zero at the bounce, so here it is the real part of the action that appears stationary with regard to $\chi.$ 

Despite this non-trivial behaviour, the WKB conditions become small again at the end of the bounce phase, and in fact the universe emerges from the bounce slightly more classically than before. This level of classicality is then maintained during the kinetic expansion phase, which is as expected as the equation of state is $w=1$ or $\epsilon = 3$ there. Overall, it is remarkable that the classicality of the universe is preserved over the course of the bounce, despite the fact that the null energy condition is violated there.

Does a blow-up of the WKB conditions signal a breakdown of the theory? This seems unlikely: as mentioned above, the Euclidean action is finite and smooth throughout, and the blow-ups are caused merely by the imaginary part of the action being momentarily stationary. A further indication that nothing singular is happening is provided by the fact that in the classical theory, linear perturbations are well-behaved through the bounce \cite{Battarra:2014tga}. Nevertheless, it is interesting to observe that the WKB conditions suggest that even a classically non-singular bounce retains a certain ``quantumness'' to it.

\paragraph{Discussion}
The results presented here provide the basis for a cosmological model devoid of singularities and infinities: the universe is finite in space and in time, and does not seem to lead to any run-away behaviour. Models built on this basis will therefore lead to predictions that are trustworthy, and that will not depend on unknowable quantities outside of the theory. This feature distinguishes the present cosmological model from existing ones. 

There are a number of foreseeable applications and improvements of the present results that are left for future work. The ghost condensate is technically the simplest model of a ghost-free non-singular bounce, but it has two drawbacks: it contains a gradient instability (whose absence likely requires the inclusion of further higher - derivative terms) and the model must be fine-tuned such that the $X$ term changes sign in an appropriate region of field space. Furthermore, the model is not entirely realistic yet: it will be desirable to include a second scalar field (for instance along the lines of \cite{Li:2013hga,Qiu:2013eoa,Fertig:2013kwa,Ijjas:2014fja}) in future studies in order to obtain realistic curvature perturbations. A particularly interesting aspect of the present study is the evolution of the WKB classicality conditions during the bounce phase. Can this evolution be understood analytically? Would it be significantly different if the gradient instability were avoided? And, most crucially, could this evolution leave an observable trace in the properties of the primordial curvature perturbations?

\paragraph{Acknowledgments}
I would like to thank Michael Koehn and Paul Steinhardt for useful comments and discussions. I gratefully acknowledge the support of the European Research Council via the Starting Grant Nr. 256994 ``StringCosmOS''.


\bibliography{NEQC_Bib}

\end{document}